\documentclass[final,3p,times,twocolumn]{elsarticle}

\usepackage[explicit]{titlesec}
\titlespacing*{\section}{0pt}{5pt}{5pt}
\titlespacing*{\subsection}{0pt}{5pt}{5pt}
\titlespacing*{\subsubsection}{0pt}{5pt}{5pt}
\usepackage{indentfirst}
\usepackage{setspace}
\usepackage{amsmath,amssymb,amsthm}
\usepackage{graphicx}
\usepackage{grffile}
\usepackage{epstopdf}
\usepackage{multirow}

\usepackage{url}
\usepackage{textcomp}
\usepackage{epstopdf}
\usepackage{algorithm}
\usepackage{algpseudocode}
\usepackage{booktabs}
\usepackage{cases}
\usepackage{xcolor}
\usepackage{caption}
\usepackage{cleveref}   
\usepackage{subfigure} 
\usepackage{float}
\usepackage{threeparttable}
\usepackage{makecell}

\def\x{\boldsymbol{x}}
\def\y{\boldsymbol{y}}

\def\W{\mathbf{W}}

\def\w{\mathbf{w}}

\def\x{\boldsymbol{x}}

\def\W{\mathbf{W}}

\def\Q{\boldsymbol{Q}}

\journal{Neural Networks}

\begin{document}

\begin{frontmatter}

\title{Quadratic Graph Attention Network (Q-GAT) for Robust Construction of Gene Regulatory Networks}




\author[neu]{Hui Zhang \corref{C2}}
\author[hit]{Xuexin An \corref{C2}}
\author[neu]{Qiang He}
\author[stevens]{Yudong Yao}
\author[uk]{Yudong Zhang}
\author[cuhk]{Feng-Lei Fan \corref{C1}}
\author[neu]{Yueyang Teng \corref{C1}}

\cortext[C2]{Co-first author.}
\cortext[C1]{Co-corresponding authors: hitfanfenglei@gmail.com (Feng-Lei Fan) and tengyy@bmie.neu.edu.cn (Yueyang Teng) }

\address[neu]{College of Medicine and Biological Information Engineering, Northeastern University, Shenyang 110169, China}
\address[hit]{School of Mathematics, Harbin Institute of Technology, Harbin 236500, China}
\address[stevens]{Department of Electrical and Computer Engineering, Stevens Institute of Technology,
 Hoboken, NJ 07030, USA}
 \address[cuhk]{Department of Mathematics, The Chinese University of Hong Kong, Hong Kong.}
\address[uk]{School of Computing and Mathematical Sciences, University of Leicester, Leicester, United Kingdom}

\begin{abstract}

Gene regulatory relationships can be abstracted as a gene regulatory network (GRN), which plays a key role in characterizing complex cellular processes and pathways. Recently, graph neural networks (GNNs), as a class of deep learning models, have emerged as a useful tool to infer gene regulatory relationships from gene expression data. However, deep learning models have been found to be vulnerable to noise, which greatly hinders the adoption of deep learning in constructing GRNs, because high noise is often unavoidable in the process of gene expression measurement. \textit{Can we preferably prototype a robust GNN for constructing GRNs?} In this paper, we give a positive answer by proposing a Quadratic Graph Attention Network (Q-GAT) with a dual attention mechanism. We study the changes in the predictive accuracy of Q-GAT and 9 state-of-the-art baselines by introducing different levels of adversarial perturbations. Experiments in the \emph{E. coli} and \emph{S. cerevisiae} datasets suggest that Q-GAT outperforms the state-of-the-art models in robustness. Lastly, we dissect why Q-GAT is robust through the signal-to-noise ratio (SNR) and interpretability analyses. The former informs that nonlinear aggregation of quadratic neurons can amplify useful signals and suppress unwanted noise, thereby facilitating robustness, while the latter reveals that Q-GAT can leverage more features in prediction thanks to the dual attention mechanism, which endows Q-GAT with the ability to confront adversarial perturbation. We have shared our code in \url{https://github.com/Minorway/Q-GAT_for_Robust_Construction_of_GRN} for readers' evaluation.
\end{abstract}

\begin{keyword}
gene expression, gene regulatory networks, graph neural networks, stability analysis
\end{keyword}

\end{frontmatter}


\section{Introduction}
\label{sec:introduction}

\indent{Gene regulatory networks (GRNs) serve as a powerful tool in cellular regulatory systems such as signal transduction and transcriptional control, which are the molecular basis of many kinds of cellular phenomena \cite{marbach_wisdom_2012, mochida_statistical_2018}. These networks are also the atlas of developmental activities that organize the spatial distribution of regulatory gene expression, explain biological processes, and prioritize candidate genes for molecular regulators in complex diseases \cite{davidson_gene_2015, marbach_wisdom_2012}. The key problem in the construction of GRNs is to infer whether a regulatory relationship exists between two genes. }


A plethora of machine learning methods have been proposed to infer gene regulatory relationships so as to establish a GRN \cite{mochida_statistical_2018}. Salleh \textsl{et al.} \cite{Salleh_multiple_2017} inferred gene regulatory relations in constructing GRNs by the multiple linear regression method after randomly extracting subgraphs from the global interaction graph. Haury \textsl{et al.} \cite{haury_tigress_2012} proposed a robust inference method by combining a stability selection criterion and the minimal angle regression. Margolin \textsl{et al.} \cite{margolin_aracne_2006} employed an information-theoretic strategy to eliminate a majority of indirect relationships identified by co-expression approaches. The gene network inference with an ensemble of trees (GENIE3) translated the regulatory network prediction problem among $p$ genes into distinct regression problems and solved them with random forests \cite{huynh-thu_inferring_2010}. The supervised inference of regulatory networks (SIRENE) proposed by Mordelet \textsl{et al.} \cite{mordelet_sirene_2008} decomposed the inference problem into many parallel binary classification problems, and every problem was addressed by a support vector machine (SVM).

Deep learning \cite{chauhan_comparison_2022, zhang_machine_2017,nguyen_machine_2019} is currently the mainstream machine learning approach. A graph neural network (GNN) is a class of deep learning models specifically dealing with graphs. Recently, many representative graph neural networks (GNNs) were applied in solving the problem of constructing GRNs \cite{bacciu_gentle_2020,li_fast_2020,spinelli_adaptive_2021,wu_comprehensive_2021}. For example, Yuan \textsl{et al.} inferred gene interactions from spatial transcriptomics data using graph convolutional networks (GCN) \cite{yuan_gcng_2020}. Wang \textsl{et al.} \cite{wang2020inductive} trained an integrated framework GRGNN with extracted subgraphs to construct GRNs. Bigness \textsl{et al.} \cite{bigness_integrating_2022} employed a GCN to predict gene expression (RNA-seq) using histone modifications and Hi-C data. The GraphReg model, developed by Karbalayghareh \textsl{et al.}, propagated local representations learned from one-dimensional inputs using graph attention network (GAT) layers over three-dimensional interaction graphs to predict gene expression (CAGE-SEQ) across genomic regions (BIN) \cite{karbalayghareh_chromatin_2022}.
Despite that these GNNs perform satisfactorily, like many commonly-used deep learning methods \cite{szegedy_intriguing_2014, akhtar_threat_2018,moosavi-dezfooli_universal_2017,finlayson_adversarial_2019,ma_understanding_2021, dai_adversarial_2018, 2020Towards,2020Adversarial}, the prediction of these GNNs are often unstable with regard to even small perturbations. Since noise is unavoidable in biological measurements, such instability greatly restricts the widespread adoption of GNNs because GNNs are likely to generate misleading interpretations of cellular phenomena, as well as the disease mechanism. Unfortunately, although the stability of GNNs is a critical issue in the construction of GRNs, to the best of our knowledge, quite a few studies focused on it, if not none.

Recently, a concept called NeuroAI \cite{zador2022toward} was coined, whose overarching idea is that a good amount of neuroscience knowledge can greatly fertilize the development of deep learning to incubate the next generation of artificial intelligence. Inspired by NeuroAI and considering that in a human brain, all kinds of intelligent behaviors are enabled by neuronal diversity, Fan \textit{et al.} introduced neuronal diversity into deep learning \cite{fan2020universal, fan2023towards} by designing quadratic neurons, which substitute the inner product in the conventional neuron with a simplified quadratic function. Quadratic networks have demonstrated the state-of-the-art performance in real-world problems, \textit{e.g.}, low-dose CT denoising \cite{fan2019quadratic}, anomaly detection \cite{liao2022heterogeneous}, and bearing fault diagnosis \cite{liao2022attention}. 

Here we propose a novel graph attention network using quadratic neurons (Q-GAT) and find that it can enhance the robustness in constructing GRNs. In addition, due to the enhanced representation power of quadratic neurons \cite{fan2020universal, fan2023one}, the predictive performance of Q-GAT is also desirable. Next, we validate the predictive ability and stability of Q-GAT in constructing GRNs by introducing adversarial perturbations to the gene expression data. The predictive ability and stability of Q-GAT are contrasted with 9 state-of-the-art models by a joint metric in \emph{E. coli} and \emph{S. cerevisiae} datasets. Experiments demonstrate the superiority of Q-GAT over its competitors in confronting adversarial perturbations. Lastly, aided by the signal-to-noise ratio (SNR) and interpretability analyses, we dissect why Q-GAT is robust. The SNR analysis informs that nonlinear aggregation of quadratic neurons can improve SNR, \textit{i.e.}, amplifying useful signals and suppressing unwanted noise, thereby facilitating robust feature extraction. The interpretability analysis reveals that Q-GAT can leverage more features than GAT in prediction, which endows Q-GAT with the ability to be less misguided by adversarial perturbation. Our contributions are threefold:

\begin{itemize}
    \item We propose Q-GAT for the task of GRNs reconstruction, which is the pioneering work in addressing the stability issue of constructing GRNs. Instead of a modification for network structures, our innovation is at the neuronal level from the perspective of neuronal diversity.

    \item Systematic experiments show that under different levels of adversarial noises, the proposed Q-GAT is superior to its competitors in terms of stability in constructing GRNs.

    \item Not just satisfied with the superior performance of Q-GAT, we move one step further to analyze why Q-GAT is more robust aided by the SNR and interpretability analyses. 

\end{itemize}

\section{Related Work}
\underline{Graph neural networks.} Node representation learning and the transmission of graph structure information are two key components of the GNN's basic structure. Node representation learning in a GNN entails providing a vector representation for each node's features. Each node can be set to start with its original feature vector. Information transfer between neighboring nodes in a graph structure refers to the GNN updating the node representation. By combining the features of its neighboring nodes, each node updates its own representation within each layer. Nodes can gradually acquire a more comprehensive and contextually relevant representation of their features after several iterations. Eventually, the node representations can be applied to graph-level tasks including graph classification, node classification, and link prediction. In order to capture the interconnections and local structure of the nodes, the graph convolutional network (GCN) was proposed to update the representation of a node by combining information about its neighbors. In order to capture the overall structure and significant characteristics of the network, graph pooling refers to the process of combining nodes or subgraphs to acquire a lower dimensional representation. 

GCN is one earlier type of GNN. Later, GAT achieves efficient node representation learning and deals with graph-level tasks by aggregating the neighbor information of nodes through an adaptive graph attention mechanism. Deep graph convolution network (DeepGCN) uses the residual/dense connectivity and dilated convolution to train very deep GNNs \cite{li_deepgcns_2019}. Graph sampling and aggregation (GraphSAGE) is a generalized induction framework that employs node feature information to construct the embeddings for previously unknown data in an efficient manner \cite{hamilton_inductive_nodate}. Autoregressive moving average filters (ARMAs) use a graph convolution layer that not only is more flexible and noise-resistant but also can retain the global graph structure better than polynomials \cite{bianchi_graph_2022}. 

Based on the GCN graph convolutional layer, deep graph convolutional neural network (DGCNN) uses a SortPooling layer to sort the graph's vertices. The SortPooling layer serves as a link between the graph convolution layer and the fully-connected layer. By recording the sorting order of the input and returning the gradient to the previous layer, the SortPooling layer not only enables the training of the previous layer but also helps learning from the topology of the global graph \cite{zhang_end--end_nodate}. Differentiable pooling (DiffPool) employs a differentiable graph pooling module to provide hierarchical graph representations, which can be integrated with diverse GNN structures \cite{ying_hierarchical_nodate}. Self-attention graph pooling (SAGPool) uses a self-attention graph pooling technique that considers both node attributes and graph topology \cite{lee_self-attention_2019}. Top-k pooling (TopKPool) employs a new graph pooling method suggested by graph U-Nets, which picks nodes to construct smaller graphs based on scalar projection values \cite{gao_graph_2021}. Graph multiset transformer (GMT) utilizes a global pooling layer based on multi-headed attention to capture interactions between nodes and their structural dependencies \cite{velickovic_graph_2018,baek_accurate_2021}. PiNet uses a generalized differentiable attention-based pooling technique combined with graph convolution operations for graph-level classification \cite{meltzer_pinet_2020}.  

Like other kinds of deep learning models, GNNs are often easily fooled by small perturbations. For example, in addition to the image-specific adversarial perturbations \cite{szegedy2013intriguing,akhtar2018threat}, Moosavi-Dezfooli \textit{et al.} \cite{moosavi2017universal} showed that “universal perturbations” can deceive a network classifier on any image. Moreover, Dai \textit{et al.} \cite{dai2018adversarial} found that GNN methods are vulnerable in both graph- and node-level classification tasks on the text and finance datasets. Thus, when applying GNNs to constructing GRNs, it is hard to tell the correctness of the inferred relationship due to the noise that emerged in biological measurements. Given the criticality of GRNs, it is highly necessary to solve the instability issue in construction.

It is noteworthy that significant advancements have been made in the reconstruction of GRNs under unlabeled conditions. First, many studies have successfully used semi-supervised learning techniques to handle both large amounts of unlabeled data and small amounts of labeled data. To better understand the underlying structure of the data, graph-based methods, for instance, can share information between labeled and unlabeled data \cite{zhu2002learning,zhou2003learning}.
The ability to share knowledge from one or more related tasks is made possible by transfer learning and multi-task learning approaches, which can be useful when dealing with scarce and unevenly distributed labeled data. Paul  \textsl{et al.}'s \cite{mignone2020multi} multi-task learning-based reconstruction of human and mouse GRNs showed that, even when the proportion of labeled exemplars to unlabeled exemplars is very small, the proposed method almost always outperforms the corresponding single task.
Graph neural networks (GNNs) have demonstrated the potential of analyzing biological data in challenging and noisy environments \cite{ma2015deep,ding2018interpretable,velickovic2018graph,gligorijevic2018deepnf}. These methods can identify subtle interactions in GRNs by automatically detecting hidden features and patterns. In the meantime, the development of integrated methods has replaced traditional ones in the reconstruction of GRNs. The robustness and accuracy of the model can be further improved by combining multiple models or strategies \cite{wang2020inductive,marbach2012wisdom,huynh2010inferring}. Additionally, causal inference has demonstrated an ever-increasing importance in this field. Causal inference can be used to explain the causal interactions between genes \cite{bar2022constrained}.

\underline{High-order units.} The investigation of high-order units can be traced back to the Group Method of Data Handling (GMDH \cite{ivakhnenko1971polynomial}), which is essentially a polynomial feature extractor:
\begin{equation}
\begin{aligned}
     Y(\x_1,\cdots,\x_n) =& a_0 + \sum_i^n a_i \x_i + \sum_i^n \sum_j^n a_{ij} \x_i \x_j \\
    & + \sum_i^n \sum_j^n \sum_k^n a_{ijk} \x_i \x_j \x_k + \cdots ,
\end{aligned}
\end{equation}
where $\x_i$ is the $i$-th input variable, and $a_i, a_{ij}, a_{ijk},\cdots$ are coefficients. Usually, GMDH only keeps the second-order terms to avoid parametric explosion for high-dimensional inputs. Recently, higher-order units were revisited \cite{chrysos2021deep, liu2021dendrite}. 
In \cite{chrysos2021deep}, the higher-order units were embedded into a deep network to reduce the complexity of the individual unit via tensor decomposition and factor sharing. Such a network achieved cutting-edge performance on several tasks. Liu and Wang \cite{liu2021dendrite} defined the so-called Gang neuron that is recursively formulated as $\mathbf{A}^{l} = \mathbf{W}^{l,l-1}\mathbf{A}^{l-1}\circ \x$ and $\mathbf{A}^0 = \x$. However, for the sake of the parameter efficiency, most polynomial neuron research concentrates on quadratic neurons, as shown in Table \ref{tab:neurons}. Among these quadratic neuron designs, Fan \textit{et al}.'s design \cite{fan2018new} is the most inclusive because other neuron designs such as \cite{jiang2020nonlinear, mantini2021cqnn, goyal2020improved, xu2022quadralib} can actually be seen as its special cases.

\begin{table}[htbp]
\setlength{\abovecaptionskip}{0.5em}
\setlength{\belowcaptionskip}{-0.5em}
\vspace{-0.2em}
\centering
\caption{An overview of the recently-proposed quadratic neurons. $\sigma(\cdot)$ denotes the nonlinear activation function. $\odot$ is the Hadamard product. Here, we omit the bias terms in a neuron. $\x \in \mathbb{R}^{n\times 1}$.
$\w_i \in \mathbb{R}^{n\times 1}$ and $\W_i \in \mathbb{R}^{n\times n}$ are the learnable weight vector and matrix, respectively.}
\scalebox{0.8}{
\begin{tabular}{|l|l|}
\hline
Work Reference           & Neuron Format          \\ \hline
\makecell[l]{Zoumpourlis \textit{et al.}\cite{zoumpourlis2017non} \\ Bu\&Karpatne \cite{bu2021quadratic} } & $\y=\sigma(\x^{\top}\W_1\x+\w_1^\top\x)$               \\ \hline
Tsapanos \textit{et al.} \cite{tsapanos2018neurons} & $\y=\sigma((\w_1^\top\x)\odot(\w_2^\top\x))$ \\
\hline
Fan \textit{et al.} \cite{fan2018new}      & $\y=\sigma((\w_1^\top\x)\odot(\w_2^\top\x)+\w_3^\top(\x\odot\x)$)        \\ \hline
\makecell[l]{Jiang \textit{et al.} \cite{jiang2020nonlinear}\\ Mantini\&Shah \cite{mantini2021cqnn}}     & {$\y=\sigma(\x^{\top}\W_1\x$)} 
                       \\ \hline
Goyal \textit{et al.} \cite{goyal2020improved}    & $\y=\sigma(\w_1^\top(\x\odot\x))$               \\ \hline
         Xu \textit{et al.} \cite{xu2022quadralib}  & {$\y=\sigma((\w_1^\top\x)\odot (\w_2^\top\x)+\w_3^\top\x)$}                             \\ \hline
\end{tabular}}
\label{tab:neurons}
\vspace{-1em}
\end{table}

\section{Preliminaries}

The complex relationships and interactions of gene regulation can be described by a GRN. Molecules like genes, proteins, transcription factors, etc. are modeled in these networks as nodes, and their interactions, regulatory relationships, etc. are connected by edges. GRNs are useful in illustrating the regulatory relationships between genes, signaling pathways, and the structure and operation of the entire regulatory system. In a transcription factor-gene network, for instance, nodes stand for transcription factors and genes, and edges signify the regulatory connections between them.

Proteins called transcription factors bind to the promoters and enhancers of target genes to control the transcriptional activity of those genes. Different transcription factors can affect the expression of target genes in a synergistic or competitive manner by binding to different regulatory regions, which typically contain specific binding sites. As a result, interactions between two different transcription factors can occur in transcription factor-gene networks, but direct interactions between two different target genes are typically absent.

Denote a collection of transcriptional factors (TFs) as $T$, a collection of target genes as $G$, and their gene expression data as $R_{i,j}$, where $i\in \{T, G \}$ and $j\in [1,n]$ for every $T$ and $G$ with $n$ arrays. 
Typically, GRNs are represented using the ${T\times(T+G)}$ matrix. The matrix has $T$ rows, which is equal to the number of transcription factors. The matrix's column number is $(T + G)$, and the first $T$ columns in the matrix represent interactions between transcription factors, while the last $G$ columns represent interactions between transcription factors and genes. This representation makes it easy to see the relationships between genes and transcription factors in GRNs, which enables us to use a linear algebraic analysis to better understand the structure and dynamics of the network. However, because this is a linear and static model, it might not fully account for the complexity of biological systems.

Our task is to obtain a GRN that is essentially an $\mathbb{R}^{T\times(T+G)}$ matrix. $E$ denotes the edges representing regulatory relationships between transcriptional factors and genes and between transcriptional factors themselves. Notably, edges between a pair of genes are prohibited because no such mechanisms are discovered so far. If $\left (i,j \right) \in E$, $A_{i,j}=1$; otherwise, $A_{i,j}=0$. 

In this study, we follow the common practice to prepare a data set to train a classifier for inference. The detailed procedure is described as follows:
\begin{enumerate}[i)]
	\item obtain candidate links from the gene expression data by thresholding the Pearson correlation coefficients;
	\item fuse node features and extract enclosed subgraphs;
	\item generate a data set based on the labeled links.

\end{enumerate}
Steps i) and ii) can significantly improve the efficiency of graph construction by eliminating many impossible links. Step iii) is essentially collecting all subgraphs. In the following, steps i) and ii) are described in detail:

\underline{Candidate links.}
When two genes have a regulatory relationship, their expression data are generally correlated, so we can use the correlation to establish candidate links. We use the Pearson correlation coefficient as a heuristic method to determine the correlation between two genes and obtain an adjacency matrix $H_{T,T+G}$. Given a threshold, the binary adjacency matrix is defined as follows:
\begin{equation}
	A_{i,j} =\begin{cases}
		1 & \text{ if  } \ H(i,j)\ge \mathrm{threshold} \\
		0 & \text{ if  } \ H(i,j)< \mathrm{threshold}.
	\end{cases}
\label{eq1}
\end{equation}
Despite greatly reducing the workload, such a primitive method must cause many pseudo links. Hence, we apply the following step to make a meticulous judgment.

\underline{Fusion of node features and extraction of subgraphs.} Embedding is a continuous feature representation in which nodes in a graph are represented as a vector by a mapping function. The learned node embeddings reflect the topology of the graph. In other words, similar graph nodes have similar low-dimensional feature patterns. This study used the Node2vec \cite{grover_node2vec_2016} method to learn the embedding features. The original gene expression data for each node reveal its biological function and contain a large amount of information. Thus, we average the gene expression data for each gene under various conditions. The average serves as the explicit features of the node which are then combined with implicit embedding features as new features.

\begin{figure}[htbp]  
 \setlength{\abovecaptionskip}{0em}
\setlength{\belowcaptionskip}{0em}
 \vspace{-1em}

\centering
\includegraphics[width=0.85\linewidth]{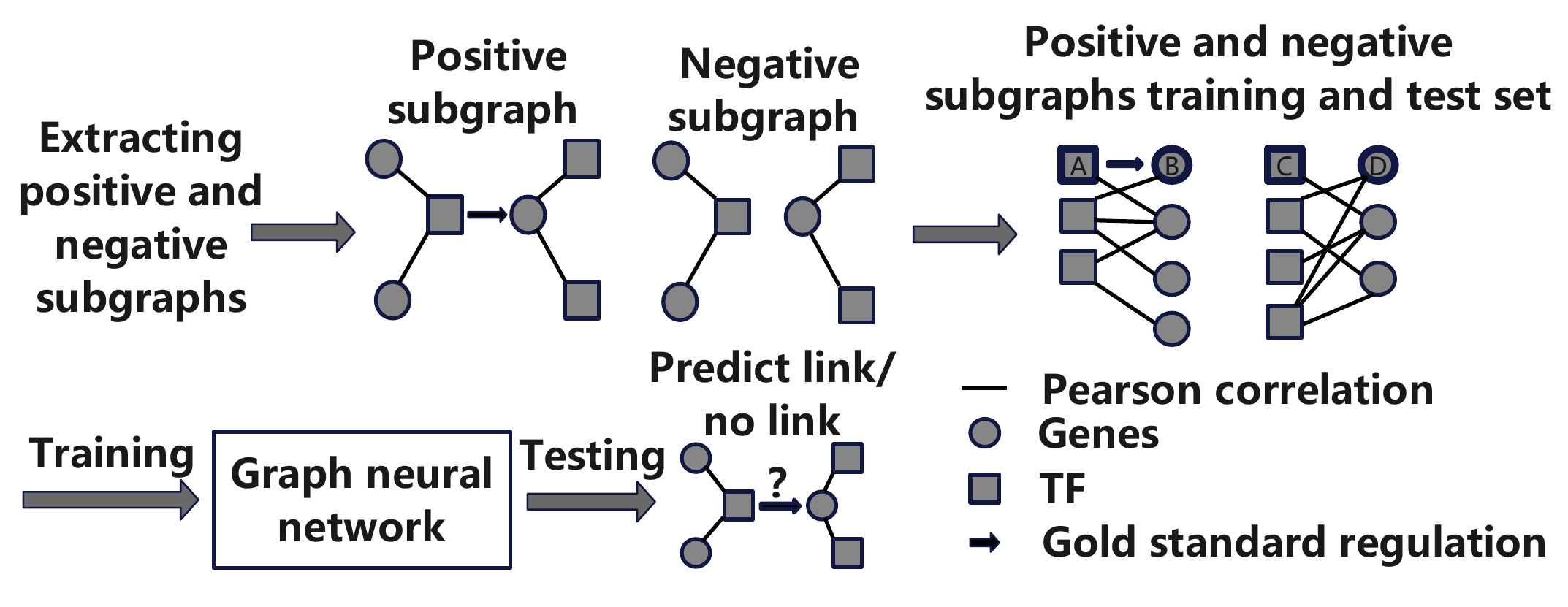} 
\caption{The procedure of extracting subgraphs for training and testing. } 
\label{fig_em}
\vspace{-1em}
\end{figure}

Next, as Figure \ref{fig_em} shows, we extract subgraphs from the above roughly constructed GRN for each linked edge, including the corresponding two nodes and their \textit{h}-hop neighbors. 
Wang \textsl{et al.} showed that adding more hops does not produce significantly better results in a GRN.
As a result, here the 1-hop subgraph is extracted. The positions of gene node pairs with links are randomly drawn on the gold standard network and divided into training and test sets. Since some regulatory relationships between genes have been predetermined by other biological tools, the corresponding subgraphs can be labeled with a link or without a link. Typically, to obtain a balanced dataset, the numbers of links and voids are cast to be approximately equal.

The final training and test sets are composed of gene node pairs with and without links. Next, the links on the Pearson correlation network corresponding to the test set positions are masked. Then, using the training and test set positions that are collected from the gold standard network, we proceed to extract subgraphs on the Pearson correlation network. A positive subgraph is extracted based on the position of the gene pair where linkage exists on the gold standard network, while a negative subgraph is extracted based on the position where linkage does not exist.

\section{Quadratic Graph Attention Network (Q-GAT)}

\begin{figure*}[htbp]  
\vspace{-0.2cm}
\centering
\includegraphics[width=1\linewidth]{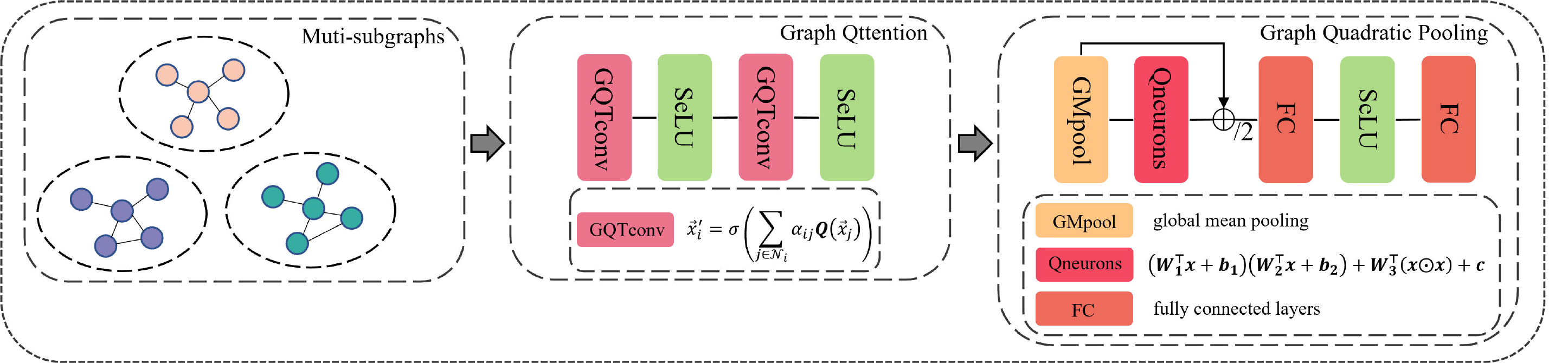} 
\caption{The structure of Q-GAT.} \label{fig1}
\vspace{-0.3cm}
\end{figure*}

\subsection{Quadratic Neuron}

Suppose that the input vector is $\boldsymbol{x} = (x_1, x_2, \ldots ,x_n)^\top$, the conventional neuron is formulated as
\begin{equation}
\sigma(l(\boldsymbol{x})) = \sigma\Big(\sum_{i=1}^{n} w_{i} x_i +b^r)\Big) =\sigma(\boldsymbol{w}^\top\boldsymbol{x}+b),
\label{lneq1}
\end{equation}
where $\sigma(\cdot)$ is a nonlinear activation function.
Encouraged by the essential role of neuronal diversity in a biological neural system, a new type of neuron called the quadratic neuron \cite{b4} was proposed to promote neuronal diversity in deep learning. A quadratic neuron \cite{b4} integrates two inner products and one power term of the input vector before nonlinear activation. Mathematically, a quadratic neuron computes
\begin{equation}
\sigma(q(\cdot)) = \sigma\Big((\boldsymbol{w}_1^\top \boldsymbol{x}+b_1)(\boldsymbol{w}_2^\top\boldsymbol{x}+b_2)+\boldsymbol{w}_3^\top(\boldsymbol{x}\odot\boldsymbol{x})+b_3\Big),
\label{qneq1}
\end{equation}
where $Q(\cdot)$ is the simplified quadratic function, $\boldsymbol{w}_1,\boldsymbol{w}_2, \boldsymbol{w}_3\in\mathbb{R}^n$ are weight vectors, and $b_1, b_2, b_3\in\mathbb{R}$ are biases. Furthermore, we can write the operation of quadratic layers as
\begin{equation}
\scalebox{0.95}{
$
\sigma(\Q(\boldsymbol{x}))
=\sigma\Big((\W_1^\top \boldsymbol{x}+\boldsymbol{b}_1)(\W_2^\top\boldsymbol{x}+\boldsymbol{b}_2)+\W_3^\top(\boldsymbol{x}\odot\boldsymbol{x})+\boldsymbol{b}_3\Big),
$}
\label{Qlayerneq1}
\end{equation}
where $\W_1,\W_2, \W_3\in\mathbb{R}^{n_1\times n_2}$ are weight matrices, and $\boldsymbol{b}_1, \boldsymbol{b}_2, \boldsymbol{b}_3 \in\mathbb{R}^{n_1}$ are bias vectors.

Quadratic networks have demonstrated the state-of-the-art performance in solving various real-world problems, \textit{e.g.}, low-dose CT denoising \cite{fan2019quadratic}, anomaly detection \cite{liao2022heterogeneous}, and bearing fault diagnosis \cite{liao2022attention}. At the same time, the superior expressive ability of quadratic neurons and networks compared to conventional ones has also been verified theoretically \cite{fan2020universal,fan2021expressivity}. These positive studies suggest that quadratic networks will be a well-performed model in GRNs.

\subsection{Qttention}
\label{sec:qttention}
The core element of the attention mechanism is to derive an importance map for the input that can notify practitioners which parts of the input are more important over others, thereby forcing the network to discriminatively leverage the input information. With such an observation, the earlier work \cite{liao2022attention} showed that a quadratic neuron can induce an attention mechanism, referred to as \textit{qttention}, by factorizing the learned quadratic function based on the canonical attention mechanism.

Mathematically, we factorize the formulation of a quadratic neuron (Eq. \eqref{qneq1}) as follows (we remove bias terms for conciseness):
\begin{equation}
\begin{aligned}
&\sigma((\boldsymbol{w}_1^\top\boldsymbol{x}+b_1)(\boldsymbol{w}_2^\top\boldsymbol{x}+b_2)+\boldsymbol{w}_3^\top(\boldsymbol{x}\odot\boldsymbol{x})+b_3) \\
=&\sigma (\boldsymbol{w}_2^\top\boldsymbol{x}( \boldsymbol{w}_1^\top\boldsymbol{x}+b_1) +b_2\boldsymbol{w}_1^\top\boldsymbol{x}+b_2b_1+\boldsymbol{w}_3^\top(\boldsymbol{x}\odot\boldsymbol{x}))\\
=&\sigma ( \boldsymbol{w}_2^{\top}( \boldsymbol{w}_1^\top\boldsymbol{x}+b_1)\boldsymbol{x}  + \boldsymbol{w}_1^\top b_2 \boldsymbol{x}  
+\left( \boldsymbol{x}\odot \boldsymbol{w}_3 \right)^{\top} \boldsymbol{x})\\
= & \sigma ( ( \underbrace{\boldsymbol{x}\odot \boldsymbol{w}_3+\boldsymbol{w}_2 (\boldsymbol{w}_1^{\top} \boldsymbol{x})}_{qttention} +\underbrace{\boldsymbol{w}_2 b_1+\boldsymbol{w}_1b_2}_{bias} )^{\top} \boldsymbol{x}) ,
\end{aligned}
\label{eq:quadratic}
\end{equation}

We exclude the bias terms $\boldsymbol{w}_2 b_1+\boldsymbol{w}_1b_2$ because they are always the same for different $\x$; therefore, they cannot work as importance scores. Then, we let
 \begin{equation}
     Qttention(\boldsymbol{x})= \boldsymbol{x}\odot \boldsymbol{w}_3+\boldsymbol{w}_2 (\boldsymbol{w}_1^{\top} \boldsymbol{x}).
\label{eq:qtt}
\vspace{-0.1em}
 \end{equation}

\subsection{Quadratic Graph Attention Network (Q-GAT) }

The GNN methods can utilize the graph structure and node features to learn a representation vector for a node or for the entire graph \cite{zhou_graph_2020}. In this study, we propose the quadratic graph attention model (Q-GAT) by exploiting quadratic neurons in a graph to process features \cite{Fan2018ANT, liao2022attention} and aggregating these features with an attention mechanism. Instead of a modification for the architecture, the key innovation of Q-GAT is the graph \textit{qttention} layer (GQT) which is a dual-attention mechanism enabled by the use of quadratic neurons, and a pooling layer based on graph qttention. Because of the nonlinearity, the use of quadratic operations not only delivers good predictive performance in constructing GRNs but also enhances the robustness of the constructed GRNs.

\subsubsection{Graph Qttention}
Based on the qttention mechanism embedded in the quadratic neuron, we propose a dual-attention graph convolution structure---Graph Qttention (GQT). As shown in Figure \ref{fig1} and Eq. \eqref{eq5}, after the input data pass through the quadratic layer, a scoring function $e: \mathbb{R}^{n} \times \mathbb{R}^{n} \rightarrow \mathbb{R}$ will act on it as follows:
\begin{equation}
e_{i j}=\operatorname{LeakyReLU}\left(\boldsymbol{a}^{\top} \cdot\left[\boldsymbol{Q} (\boldsymbol{x}_{i}) \| \boldsymbol{Q}(\boldsymbol{x}_{j})\right]\right)
\label{eq5}
\end{equation}
where $\boldsymbol{a}^{\top}\in \mathbb{R}^{2n }$ is learned,  $\|$ is the concatenation operation, the attention coefficient $e_{ij}$ reflects the significance of $\x_j$ to $\x_i$, and $\mathcal{N}_{i}$ is a collection of neighbors of the node $i$. In our model, only $e_{i j}$ for nodes $j \in \mathcal{N}_{i}$ is computed; therefore, more computing resources are saved than directly calculating the relative coefficients among all nodes. Lastly, the softmax function is applied to normalize $e_{i j}$ into $[0,1]$:
\begin{equation}
\alpha_{i j} = \operatorname{softmax}_{j}\left(e_{i j}\right) = \frac{\exp \left(e_{i j}\right)}{\sum_{k \in \mathcal{N}_{i}} \exp \left(e_{i k}\right)}
\label{eq6}
\end{equation}

\textit{Why is the proposed GQT a dual attention mechanism?} First, the importance of each output node can be represented by its neighbors with the obtained attention coefficient $\alpha_{i j}$. On the other hand, the learned quadratic mapping $\boldsymbol{Q}$ enjoys the qttention mechanism which can signify the importance of features of a given node.

Furthermore, multi-head attention can be also explored. We implement K-independent dual-attention mechanisms of Eq. \eqref{eq7} and then concatenate the resulting features in Eq. \eqref{eq8}:
\begin{equation}
\x_{i}^{\prime} = \sigma\left(\sum_{j \in \mathcal{N}_{i}} \alpha_{ij} \boldsymbol{Q}(\x_{j})\right)
\label{eq7}
\end{equation}

\begin{equation}
\x_{i}^{\prime}=\overset{K}{\underset{k=1}{\|}} \sigma\left(\sum_{j \in \mathcal{N}_{i}} \alpha_{i j}^{k} \boldsymbol{Q}^{K}(\x_{j})\right)
\label{eq8}
\end{equation}
where $\alpha_{i j}^{k}$ and $\boldsymbol{Q}^{K}$ are the K-term attention coefficient and qttention mapping, respectively.

\subsubsection{Graph Quadratic Pooling}
We design the graph quadratic pooling layer, which highlights the use of quadratic operations. The reason is to analogize an attention-based pooling layer. Global mean pooling is generally employed to obtain the global graph representation. Although taking the average or maximum as the pooling is directly feasible, it ignores the node difference. In Subsection \ref{sec:qttention}, we discuss that quadratic neurons can assign different weights to nodes through diverse feature weights. Therefore, graph quadratic pooling is an attention-based pooling when the weights of quadratic neurons are updated.

First, assume that the output of the graph qttention layer is $\x$, we apply global mean pooling $\boldsymbol{GM}$ to obtain a preliminary graph representation without considering node differences:
\begin{equation}
\boldsymbol{x_{GM}}=\boldsymbol{GM(\x)}.
\label{eq9}
\end{equation}
Further, the graph quadratic pooling can be described as Eq. \eqref{eq10}: 
\begin{equation}
\boldsymbol{x_{GQ}}=\frac{1}{2}(\boldsymbol{x_{GM}}+\boldsymbol{Q}(\boldsymbol{x_{GM}})),
\label{eq10}
\end{equation}
where $\boldsymbol{Q}$ denotes the quadratic function.

Due to the attention mechanism of quadratic neurons, Eq. \eqref{eq10} allows different nodes to be assigned with different weights, thus dynamically regulating the relationship between nodes. Moreover, keeping item $\boldsymbol{x_{GM}}$ is helpful to improve the stability of the network and facilitate the optimization process. The attention mechanism of quadratic neurons is indeed powerful. In this article, however, discussing the attention mechanism of graph quadratic pooling is not the main focus. It will be used to enhance the robustness of the model. Furthermore, considering that GNNs are vulnerable to noise, we manually add randomness in the form of quadratic neurons with certain initialization conditions in the pooling layer, which can further improve the robustness. 


\section{Experiments}

\subsection{Datasets}

We study the predictive accuracy and robustness of our method and 9 powerful GNN counterparts using the public \emph{E. coli} and \emph{S. cerevisiae} datasets from the DREAM5 challenge \cite{marbach_wisdom_2012}. This dataset is a benchmark that contains a set of known regulatory interactions. The detailed statistics are listed in \Cref{tbl1}. We split data into the training and test sets, with a ratio of 7:3. The threshold for constructing the initial GRN is 0.8. The embedding feature is a one-dimensional vector. 

\begin{table}[htbp]
\setlength{\abovecaptionskip}{0.5em}
\setlength{\belowcaptionskip}{-0.5em}
\vspace{-0.2em}
\centering
\caption{The statistics of the used datasets.}
 \scalebox{0.8}{
	\begin{tabular}{|l|c|c|c|c|}
		\hline
		Species   & TFs & Target Genes &  Links & Samples\\ \hline
		\emph{E. coli}         & 344 & 4177    & 2066  & 805     \\ \hline 
		\emph{S. cerevisiae}  & 333 & 5617    & 3940  & 536     \\ \hline
	\end{tabular}}
\label{tbl1}
\vspace{-1em}
\end{table}

\subsection{Adversarial Perturbations}

Stability refers to that a model to what extent is robust to adversarial perturbations. Despite that one can add perturbations to the dataset during the training stage, which is called data poisoning \cite{miller_adversarial_2020, chen_survey_2020}, it is common to introduce perturbations into the test set after the model is already trained over the training set.  

\begin{table*}[htbp] 
\setlength{\abovecaptionskip}{0.5em}
\setlength{\belowcaptionskip}{-0.5em}
\vspace{-0.2em}
	\centering
	\setlength{\tabcolsep}{12pt}
	\caption{Performance of GNNs with multi-perturbations under different levels for \emph{E. coli} datasets (mean$\pm$ deviation).}\label{tbl2}
	 \scalebox{0.8}{
 \begin{tabular}{lccccccc}
		\toprule
		Methods & $c=$0.0         & $c=$0.05        & $c=$0.1         & $c=$0.15        & $c=$0.2         & $c=$0.25        & $c=$0.3    \\ 
		\midrule
		DGCNN              & 0.735 & 0.707$\pm$0.040  & 0.693$\pm$0.044 & 0.705$\pm$0.042 & 0.677$\pm$0.041 & 0.662$\pm$0.032 & 0.666$\pm$0.038 \\
		DiffPool           & 0.637 & 0.599$\pm$0.063 & 0.575$\pm$0.067 & 0.593$\pm$0.064 & 0.547$\pm$0.062 & 0.528$\pm$0.046 & 0.536$\pm$0.054 \\
		GMT                & 0.718 & 0.706$\pm$0.026 & 0.689$\pm$0.032 & 0.698$\pm$0.030  & 0.680$\pm$0.026  & 0.673$\pm$0.022 & 0.671$\pm$0.024 \\ 
		PiNet              & \textbf{0.803} & 0.662$\pm$0.184 & 0.622$\pm$0.191 & 0.674$\pm$0.179 & 0.513$\pm$0.171 & 0.460$\pm$0.134  & 0.496$\pm$0.165 \\ 
		ARMA               & 0.704 & 0.679$\pm$0.044 & 0.663$\pm$0.048 & 0.674$\pm$0.048 & 0.637$\pm$0.048 & 0.620$\pm$0.039  & 0.620$\pm$0.043  \\ 
		DeepGCN            & 0.748 & 0.674$\pm$0.114 & 0.633$\pm$0.124 & 0.678$\pm$0.116 & 0.570$\pm$0.116  & 0.535$\pm$0.089 & 0.554$\pm$0.108 \\ 
		SAGPool            & 0.650 & 0.637$\pm$0.046 & 0.613$\pm$0.056 & 0.622$\pm$0.053 & 0.590$\pm$0.051  & 0.570$\pm$0.038  & 0.565$\pm$0.043 \\
		TopKPool           & 0.782 & 0.660$\pm$0.184  & 0.578$\pm$0.201 & 0.630$\pm$0.195  & 0.465$\pm$0.172 & 0.439$\pm$0.151 & 0.443$\pm$0.151 \\ 
		GraphSAGE          & 0.773 & 0.664$\pm$0.161 & 0.620$\pm$0.172  & 0.667$\pm$0.162 & 0.515$\pm$0.147 & 0.479$\pm$0.119 & 0.477$\pm$0.118 \\ 
		Q-GAT(\textbf{ours})          & 0.726 & \textbf{0.707$\pm$0.020} & \textbf{0.700$\pm$0.024}  & \textbf{0.710$\pm$0.022} & \textbf{0.691$\pm$0.023} & \textbf{0.681$\pm$0.020} & \textbf{0.679$\pm$0.021} \\ 
		\bottomrule
	\end{tabular}}
\vspace{-1em}
\end{table*}

\begin{table*}[ht] 
\setlength{\abovecaptionskip}{0.5em}
\setlength{\belowcaptionskip}{-0.5em}
\vspace{1em}
	\centering
	\setlength{\tabcolsep}{12pt}
	\caption{Performance of early-stopping GNNs and Q-GAT with multi-perturbations under different levels for \emph{E. coli} datasets (mean$\pm$ deviation).}\label{tbl4}
 	 \scalebox{0.8}{
	\begin{tabular}{lccccccc}
		\toprule
		Methods & $c=$0.0         & $c=$0.05        & $c=$0.1         & $c=$0.15        & $c=$0.2         & $c=$0.25        & $c=$0.3    \\ 
		\midrule
		PiNet              & 0.726 & 0.679$\pm$0.072 & 0.653$\pm$0.078 & 0.676$\pm$0.075 & 0.614$\pm$0.071 & 0.600$\pm$0.061  & 0.611$\pm$0.069 \\ 
		TopKPool           & 0.693 & 0.646$\pm$0.071  & 0.614$\pm$0.084 & 0.631$\pm$0.082  & 0.567$\pm$0.077 & 0.543$\pm$0.061 & 0.542$\pm$0.060	 \\ 
		Q-GAT(\textbf{ours})          & 0.726 & \textbf{0.707$\pm$0.020} & \textbf{0.700$\pm$0.024}  & \textbf{0.710$\pm$0.022} & \textbf{0.691$\pm$0.023} & \textbf{0.681$\pm$0.020} & \textbf{0.679$\pm$0.021} \\ 
		\bottomrule
	\end{tabular}}
\vspace{-1em}
\end{table*}

We add Gaussian perturbation to data as follows:
\begin{equation}
	z^{*}=z+ \varepsilon\label{eq:perturbation}
\end{equation}
where $z$ and $z^{*}$ represent the original and perturbed data, respectively, and $\varepsilon \sim N\left ( 0,\sigma ^{2}  \right )$. As well known, $P\left \{ \left | \varepsilon \right |\le 3\sigma   \right \} =0.997\approx 1$; so we denote $c=3\sigma$ as the absolute upper error limit.

We add different levels of perturbations by casting $c=$ 0.05, 0.1, 0.15, 0.2, 0.25, and 0.3, respectively, in Eq. \eqref{eq:perturbation}. We repeat the above steps to obtain the perturbed gene pairs whose labels will be predicted by our Q-GAT and other GNN methods. All methods are tested on 30 independent perturbations. Then, the mean score and its standard deviation are analyzed.

\begin{figure}[ht]
\setlength{\abovecaptionskip}{0.1cm}
\setlength{\belowcaptionskip}{-0.0cm}
 \vspace{-0.2cm}

    \centering
    \subfigure[\emph{E. coli}]{
    \includegraphics[width=3in]{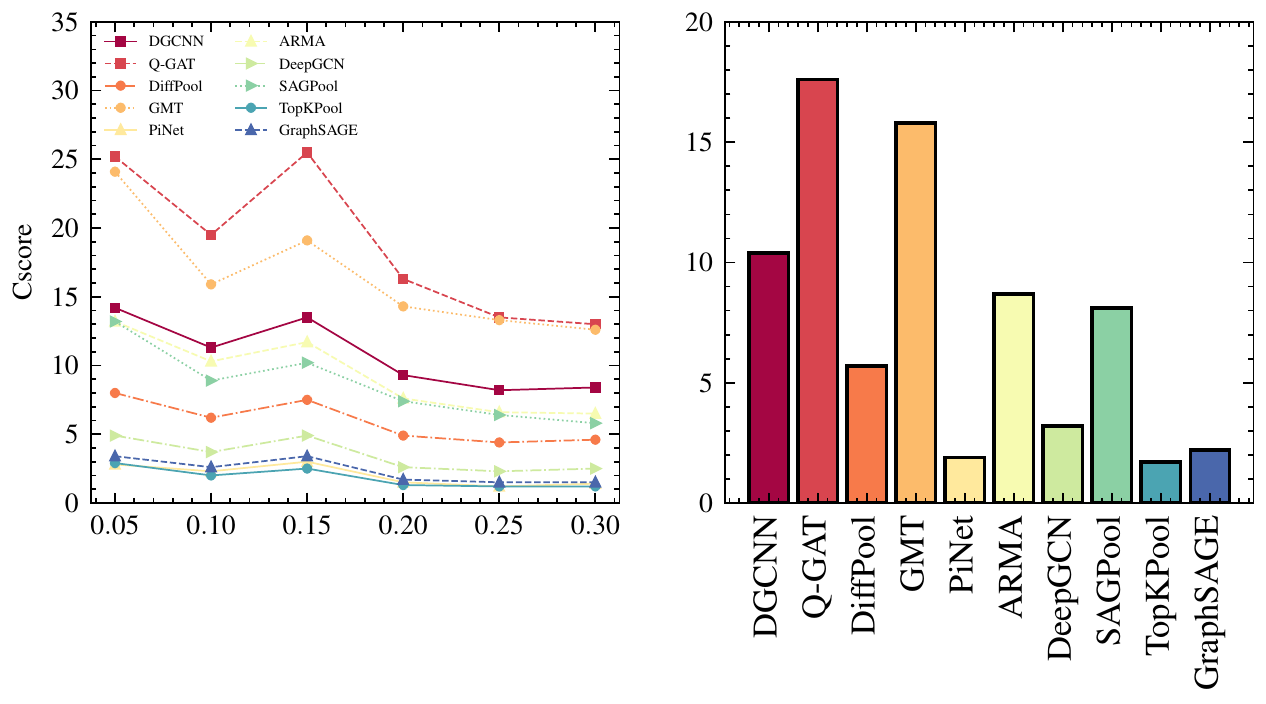} 
    }
    \subfigure[\emph{S. cerevisiae}]{
    \includegraphics[width=3in]{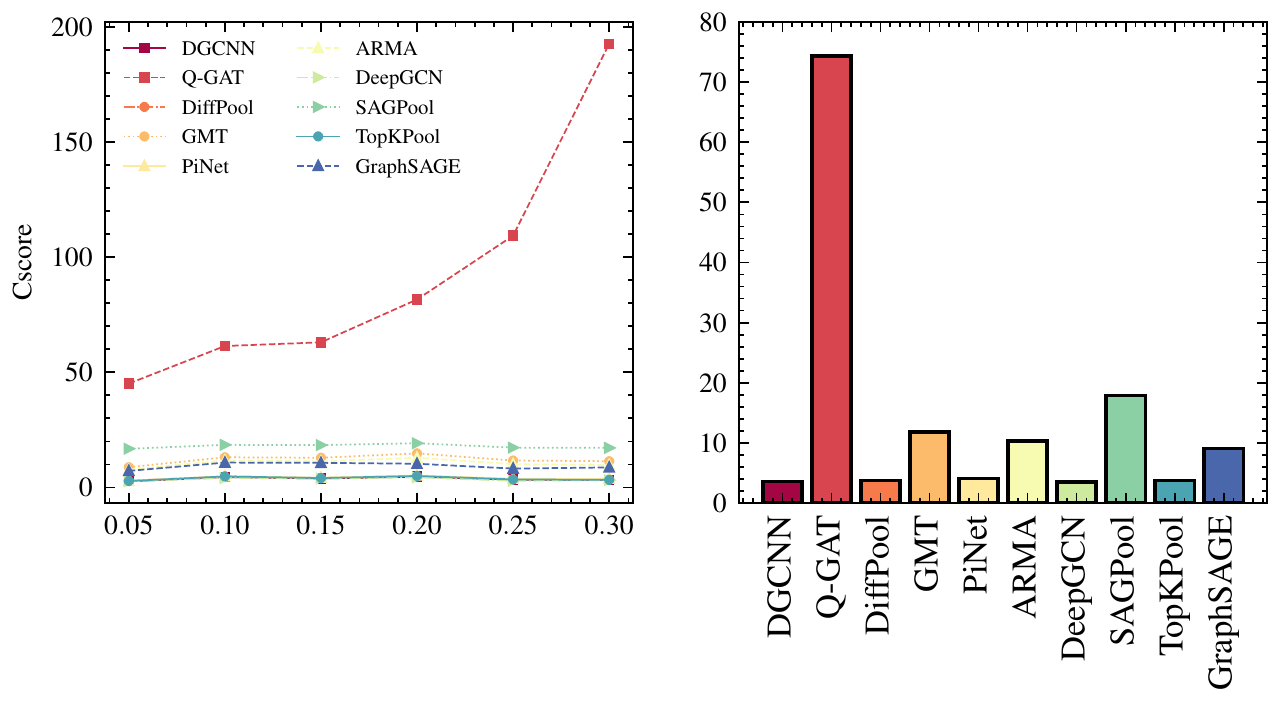} 
    }
    \DeclareGraphicsExtensions.
    \caption{(a) Cscores at different perturbation levels on the left and total cscores for all the methods on the right on the \emph{E. coli} datasets; (b) Cscores at different perturbation levels on the left and total cscores for all the methods on the right on the \emph{S. cerevisiae} datasets.}
    \vspace{-0.4cm}
    \label{fig2}
    \end{figure}

\subsection{Evaluation Metrics}

Although the focus of this draft is the stability, the predictive performance should be also cared for the usefulness of the model. To evaluate both the stability and predictive performance of different models in constructing GRNs, we propose a new and a more balanced evaluation metric as follows:
	\begin{equation}
		\mathrm{Cscore}=\frac{\mathrm{Mean}}{\mathrm{Bias}}, \label{eq3}
	\end{equation}
    where		
	\begin{equation}
		\mathrm{Bias}=\sqrt{\frac{ {\textstyle \sum_{i=1}^{N}} {{(I_{i}-\overline{I}})}^{2}}{N-1}}, \label{eq2}
	\end{equation}
	 and $\mathrm{Mean}$ is the average accuracy of $N$ repeated experiments at the same perturbation level, $\overline{I}$ is the accuracy for the perturbation-free dataset, and $I_{i}$ denotes the accuracy for the $i$-th perturbed dataset. A larger mean and a smaller bias lead to a larger Cscore. Note that one can use the variability (mean divided by standard deviation).
We do not use this because it will become very large when the standard deviation is small. However, at this time, the mean could be very low, which means that the model still cannot be used.

\subsection{Stability Results}

\begin{table*}[htbp]  
\setlength{\abovecaptionskip}{0.5em}
\setlength{\belowcaptionskip}{-0.5em}
\vspace{-0.2em}
	\centering
	\setlength{\tabcolsep}{12pt}
	\caption{The hyperparameter settings, training iterations, and loss functions applied for various models.}\label{tbl5}
 	 \scalebox{0.8}{
	 \begin{tabular}{lccccccc}
		\toprule
		Method & Learning rate  & Hidden layers   & Iterations on \emph{E. coli} & Iterations On \emph{S. cerevisiae}  & Loss function  \\ 
		\midrule
		DGCNN              & $10^{-4}$  & {[}32,32{]}        & 100 & 150 & cross-entropy loss  \\
		DiffPool           & $10^{-6}$  & {[}64,64{]}$\times$3 & 50 & 50 & cross-entropy loss   \\
		GMT                & $10^{-4}$  & {[}32,32{]}        & 150 & 150 & cross-entropy loss \\
		PiNet              & $10^{-4}$  & {[}64,64{]}        & 150 & 150 & cross-entropy loss  \\
		ARMA               & $10^{-4}$  & {[}128{]}          & 100 & 150 & cross-entropy loss  \\ 
		DeepGCN            & $10^{-4}$  & {[}64{]}$\times$28   & 100 & 150 & cross-entropy loss  \\ 
		SAGPool            & $10^{-4}$  & {[}128,128,128{]}  & 150 & 150 & cross-entropy loss  \\ 
		TopKPool           & $10^{-4}$  & {[}128,128,128,64{]} & 100 & 100 & cross-entropy loss \\
		GraphSAGE 		   & $10^{-4}$  & {[}32,32,32{]}     & 100 & 150 & cross-entropy loss  \\
		Q-GAT(\textbf{ours})& $10^{-5}$ & {[}256,256,128{]} & 150  & 150 & cross-entropy loss  \\ 
		\bottomrule
 	\end{tabular}}
\vspace{-1em}
\end{table*}

Table \ref{tbl5} displays the hyperparameter settings, training iterations, and loss functions applied for various models. In order to achieve the best performance for each model, the number of iterations and learning rate are adjusted in this paper based on the model convergence during the model-specific training process.

The comparative results for \emph{E. coli} and \emph{S. cerevisiae} are summarized in Tables \ref{tbl2} and \ref{tbl3}, respectively. When there are no perturbations in the \emph{E. coli} dataset, we find that PiNet has the highest accuracy (0.803) among all models, and DiffPool has the lowest one (0.637). The predictive scores of DGCNN, GMT, ARMA, and Q-GAT are very close, which are 0.735, 0.718, 0.704, and 0.726, respectively. The accuracy of all models drops after perturbation. But DiffPool, PiNet, DeepGCN, SAGPool, TopKPool, and GraphSAGE's accuracy dramatically drops as the perturbation level increases, while Q-GAT and GMT effectively confront the disturbance. Relative to GMT, Q-GAT achieves the highest score at each and every level of disturbance, and the gap is substantial when $c=0.1$ and $0.15$, suggesting that Q-GAT has stronger stability.

\begin{table*}[htbp]  
\setlength{\abovecaptionskip}{0.5em}
\setlength{\belowcaptionskip}{-0.5em}
\vspace{1em}
	\centering
	\setlength{\tabcolsep}{12pt}
	\caption{Performance of GNNs with multi-perturbations under different levels for \emph{S. cerevisiae} datasets (mean$\pm$ deviation).}\label{tbl3}
 	 \scalebox{0.8}{
	\begin{tabular}{lccccccc}
		\toprule
		Method & $c=$0.0         & $c=$0.05        & $c=$0.1         & $c=$0.15        & $c=$0.2         & $c=$0.25        & $c=$0.3   \\ 
		\midrule
		DGCNN              & 0.765 & 0.584$\pm$0.115 & 0.675$\pm$0.119 & 0.657$\pm$0.125 & 0.685$\pm$0.117 & 0.635$\pm$0.129 & 0.629$\pm$0.132 \\
		DiffPool           & 0.726 & 0.556$\pm$0.082 & 0.626$\pm$0.081 & 0.609$\pm$0.087 & 0.628$\pm$0.083 & 0.587$\pm$0.090  & 0.587$\pm$0.090  \\
		GMT                & 0.767 & 0.698$\pm$0.038 & 0.728$\pm$0.038 & 0.726$\pm$0.037 & 0.735$\pm$0.036 & 0.721$\pm$0.040  & 0.721$\pm$0.041 \\
		PiNet              & 0.756 & 0.587$\pm$0.100 & 0.668$\pm$0.104 & 0.653$\pm$0.108 & 0.676$\pm$0.101 & 0.633$\pm$0.108 & 0.635$\pm$0.108 \\
		ARMA               & 0.762 & 0.691$\pm$0.050 & 0.727$\pm$0.050 & 0.724$\pm$0.051 & 0.733$\pm$0.048 & 0.715$\pm$0.051 & 0.712$\pm$0.052 \\ 
		DeepGCN            & 0.761 & 0.587$\pm$0.123 & 0.666$\pm$0.126 & 0.658$\pm$0.129 & 0.676$\pm$0.125 & 0.606$\pm$0.130 & 0.627$\pm$0.136 \\ 
		SAGPool            & 0.598 & 0.566$\pm$0.009 & 0.569$\pm$0.009 & 0.569$\pm$0.009 & 0.570$\pm$0.008 & 0.567$\pm$0.008 & 0.567$\pm$0.008 \\ 
		TopKPool           & 0.765 & 0.592$\pm$0.113 & 0.680$\pm$0.110 & 0.659$\pm$0.117 & 0.685$\pm$0.109 & 0.633$\pm$0.118 & 0.621$\pm$0.117 \\
		GraphSAGE 		   & \textbf{0.775} & 0.693$\pm$0.046 & 0.730$\pm$0.049 & 0.729$\pm$0.048 & 0.727$\pm$0.050 & 0.708$\pm$0.052 & 0.714$\pm$0.053 \\
		Q-GAT(\textbf{ours})          & 0.765 & \textbf{0.749$\pm$0.004} & \textbf{0.754$\pm$0.004}  & \textbf{0.754$\pm$0.004} & \textbf{0.757$\pm$0.004} & \textbf{0.759$\pm$0.003} & \textbf{0.764$\pm$0.004} \\ 
		\bottomrule
 	\end{tabular}}
\vspace{-1em}
\end{table*}

It is observed that although PiNet and TopKPool have achieved the highest and the second-highest accuracy in the original test set, respectively, they do not perform well even under very small perturbations. To examine if this is because of overfitting, the early stopping strategy is applied to PiNet and TopKPool. \Cref{tbl4} shows that the test results of PiNet and TopKPool under the early stopping on the \emph{E. coli} dataset. We control the magnitude of the early stopping to make the prediction accuracy of PiNet and TopKPool similar to that of Q-GAT when there is no perturbation. It can be seen that after the early-stopping is applied, the prediction accuracy of PiNet and TopKPool increases at all perturbation levels, compared to the performance of their full convergence. However, PiNet and TopKPool are still inferior to Q-GAT at every disturbance level.

For the \emph{S. cerevisiae} dataset, GraphSAGE and SAGPool obtain the highest and lowest accuracy (0.775 and 0.598) under no perturbation, respectively, while Q-GAT achieves the accuracy of 0.765, close to that of GraphSAGE. Under different levels of perturbations, we observe that Q-GAT's accuracy is always the top, and its standard deviation is the lowest, which shows strong stability. Moreover, the accuracy of Q-GAT only decreases moderately when the perturbation level goes up. In contrast, the accuracy of all other methods exhibits a relatively large degradation, and their standard deviations are also generally larger than Q-GAT. For example, when $c=0.3$, Q-GAT has the highest accuracy of 0.764, while SAGPool has the lowest accuracy of 0.567. Q-GAT performs significantly better than SAGPool.

\Cref{fig2} illustrates Cscores of all models over two datasets with respect to different perturbation levels. For the \emph{E. coli} dataset, as the perturbation rises, holistically, Cscores of all methods exhibit a downward trend. But when $c=$ 0.15, for each method, the Cscores of different models slightly increase. Q-GAT has the highest Cscores across all perturbation levels. For the \emph{S. cerevisiae} dataset, Q-GAT's Cscores are always substantially higher than those obtained using other methods, as the perturbation level increases. Specifically, the Cscores at different perturbation levels for the \emph{E. coli} and \emph{S. cerevisiae} dataset are calculated as shown in Figure \ref{fig2}. For the \emph{E. coli} dataset, the results indicate that Q-GAT, with a Cscore of 17.6, is the best method, whereas DGCNN, ARMA, and SAGPool have Cscores of 10.4, 8.7, and 8.1, respectively, which are not significantly different. For the \emph{S. cerevisiae}, Q-GAT, with a Cscore of 74.3, is much higher than other methods.


\begin{table}[htbp]
\setlength{\abovecaptionskip}{0.5em}
\setlength{\belowcaptionskip}{-0.5em}
\vspace{-0.2em}
\centering
\caption{The similarity of prediction labels of GMT and Q-GAT in the \emph{S. cerevisiae} datasets between $c = 0.15$ or $c=0.3$ and $c = 0$.}
\scalebox{0.8}{
\begin{tabular}{|l|c|c|}
\hline
Similarity & $(c = 0.15,c=0)$ & $(c = 0.3,c=0)$              \\ \hline
GMT & 0.7216 & 0.7288  \\ \hline
Q-GAT & 0.9148 & 0.9013      \\ \hline
\end{tabular}}
\label{tab:similarity}
\vspace{-1em}
\end{table}

Although it is excellent to maintain high accuracy when the perturbation level goes up, the reason may be that the original prediction labels are misclassified when no perturbation is added, but are correctly classified when the perturbation is added. For a stable construction, we hope to ensure that the classification results under different perturbation levels are as close to the original prediction under no perturbation as possible. Thus, it means that the method can effectively suppress the interference of noise.

To describe such an issue, a similarity index is introduced here. Let us firstly predefine a function $f(x,y)$
\begin{equation}
	f(x,y) =\begin{cases}
		1 , x=y \\
		0 , x\neq y.
	\end{cases}
\end{equation}

Then, the similarity index can be defined as
\begin{equation}
\mathrm{Sim}(\boldsymbol{x},\boldsymbol{y})=\frac{\sum_{i=1}^{N}{f(x_i,y_i)}}{N},
\end{equation}
where $\boldsymbol{x}=\left\{x_1,\ldots,x_N\right\}$ and $\boldsymbol{y}=\left\{y_1,\ldots,y_N\right\}$ denote prediction labels of a model like Q-GAT at two different noise levels, and $N$ denotes sequence length.

Here, the similarity of prediction labels of GMT and Q-GAT in the \emph{S. cerevisiae} datasets between $c = 0.15$ or $c =0.3$ and $c = 0$ are calculated to compare if GMT and Q-GAT can constrain the interference of noise well. The reason for selecting GMT and Q-GAT is GMT and Q-GAT are the top performers in terms of stability. Table \ref{tab:similarity} shows that when the noise is stronger, the prediction results of Q-GAT are still highly consistent with that of Q-GAT in the absence of noise, even when $c=0.3$, but the results of GMT are significantly different from that of GMT in the absence of noise. It indicates that Q-GAT not only better suppresses the influence of noise but also is more stable than GMT.

\subsection{Visualization}
To better understand the representation ability of our model, we use t-SNE to visualize features generated by intermediate layers of Q-GAT and GMT. The t-SNE \cite{maaten_visualizing_2008} is a widely-used tool to visualize high-dimensional data by dimension reduction. Figure \ref{fig4} shows the visualization of Q-GAT and GMT in the \textit{S. cerevisiae} dataset at the perturbation level of $c = 0.3$. Before passing through the convolution layer, the features of two models are aliased. When features are processed in the convolution layer, they are coarsely classified. As can be seen, in our Q-GAT, features of the same category start to gather, while features in GMT are still heavily intermixed. In our Q-GAT, features of the same category do not gather into one cluster. Instead, they form into multiple clusters, which might be a shortcut path for classification because compressing all features into a cluster is harder. When features pass through the pooling layer, they are sufficiently classified. In our Q-GAT, features of two classes are clearly separated, while features generated by GMT are still aliased, which demonstrates that Q-GAT has better representation ability.

\subsection{Ablation Study}
Although graph \textit{qttention} (GQT) and graph quadratic pooling (GQP) are excellent predictors when combined, it is difficult to know how much they can contribute to the whole model individually. To make this clear, we conduct ablation experiments on \emph{S. cerevisiae} dataset. By replacing all quadratic neurons in GQT with conventional neurons, we obtain a model called Q-GAT (no GQT). This can be considered a degraded version of Q-GAT, known as GAT. On the other hand, Q-GAT (no GQP) is obtained by replacing GQP with global mean pooling.

\begin{table*}[htbp] 
\setlength{\abovecaptionskip}{0.2cm}
\setlength{\belowcaptionskip}{-0.0cm}
	\centering
	\setlength{\tabcolsep}{12pt}
	\caption{Ablation results of Q-GAT under different levels of perturbations for \emph{S. cerevisiae} datasets (mean$\pm$ deviation).}\label{tbl6}
 	 \scalebox{0.75}{
	\begin{tabular}{lccccccc}
		\toprule
		Methods & $c=$0.0         & $c=$0.05        & $c=$0.1         & $c=$0.15        & $c=$0.2         & $c=$0.25        & $c=$0.3    \\ 
		\midrule
		Q-GAT(\textbf{no GQT})              & 0.745 & 0.576$\pm$0.116 & 0.661$\pm$0.122 & 0.669$\pm$0.120 & 0.679$\pm$0.117 & 0.638$\pm$0.129  & 0.623$\pm$0.132 \\ 
		Q-GAT(\textbf{no GQP})           & \textbf{0.769} & 0.732$\pm$0.011 & 0.743$\pm$0.011 & 0.740$\pm$0.012  & 0.745$\pm$0.011 & 0.744$\pm$0.012 & 0.746$\pm$0.013	 \\ 
		Q-GAT         & 0.765 & \textbf{0.749$\pm$0.004} & \textbf{0.754$\pm$0.004}  & \textbf{0.754$\pm$0.004} & \textbf{0.757$\pm$0.004} & \textbf{0.759$\pm$0.003} & \textbf{0.764$\pm$0.004} \\ 
		\bottomrule
	\end{tabular}}
 \vspace{-0.3cm}
\end{table*}


\begin{figure}[ht!]  
\setlength{\abovecaptionskip}{0cm}
\setlength{\belowcaptionskip}{-0.3cm}
		\centering
  
           
		\subfigure[Before convolution]{
			\begin{minipage}[b]{0.2\textwidth}
				\includegraphics[width=1\textwidth]{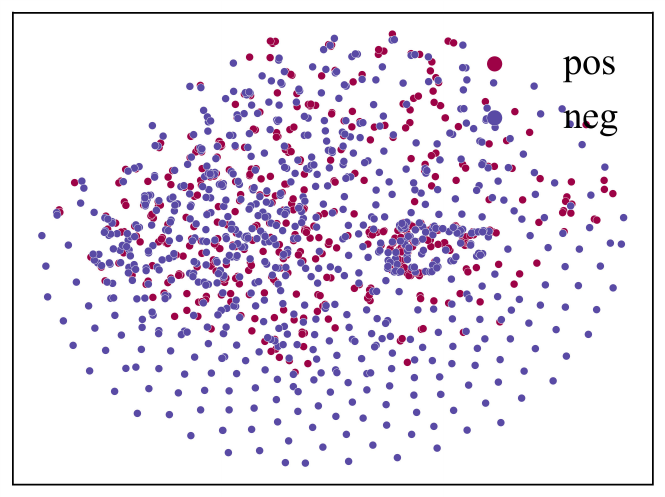} 
				\includegraphics[width=1\textwidth]{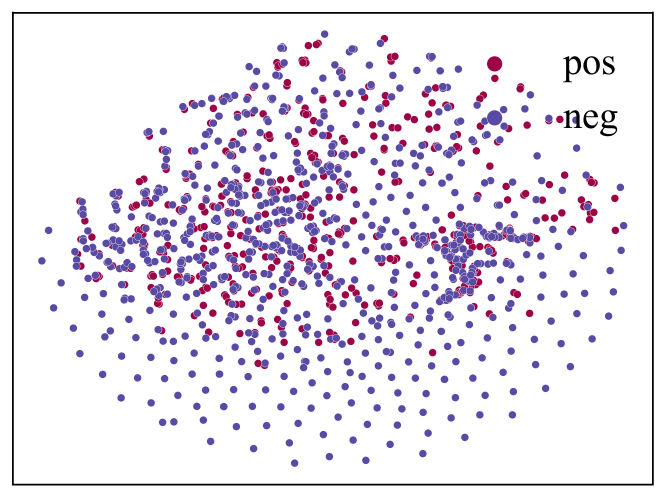}
			\end{minipage}
		}
		\subfigure[Before pooling]{
			\begin{minipage}[b]{0.2\textwidth}
				\includegraphics[width=1\textwidth]{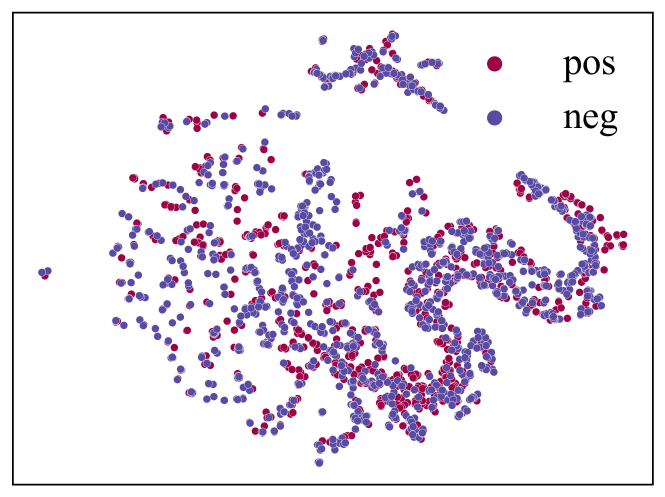} 
				\includegraphics[width=1\textwidth]{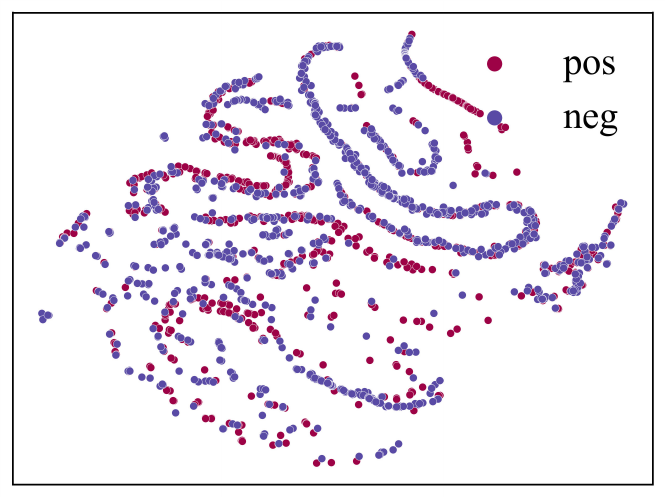}
			\end{minipage}
		}
		\subfigure[After pooling]{
			\begin{minipage}[b]{0.2\textwidth}
				\includegraphics[width=1\textwidth]{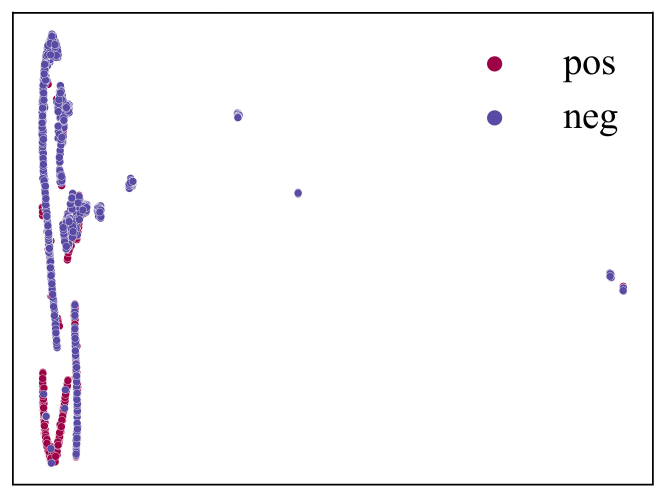} 
				\includegraphics[width=1\textwidth]{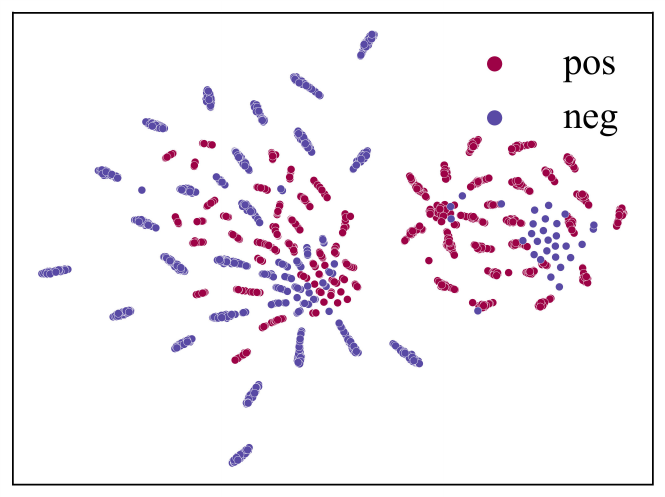}
			\end{minipage} 
		} 		
		\subfigure[Output features]{
			\begin{minipage}[b]{0.2\textwidth}
				\includegraphics[width=1\textwidth]{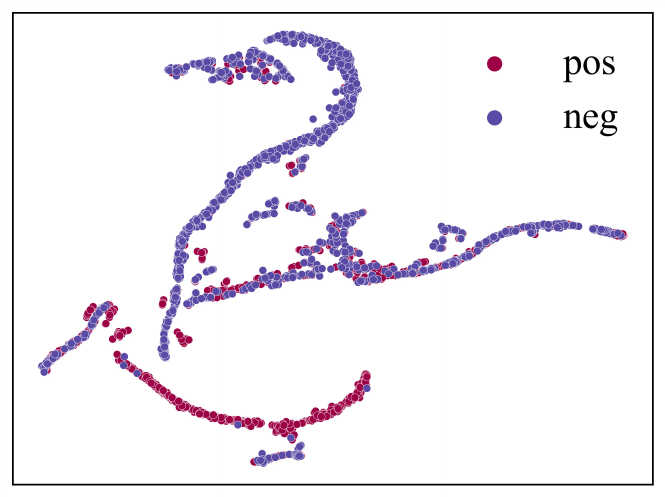} 
				\includegraphics[width=1\textwidth]{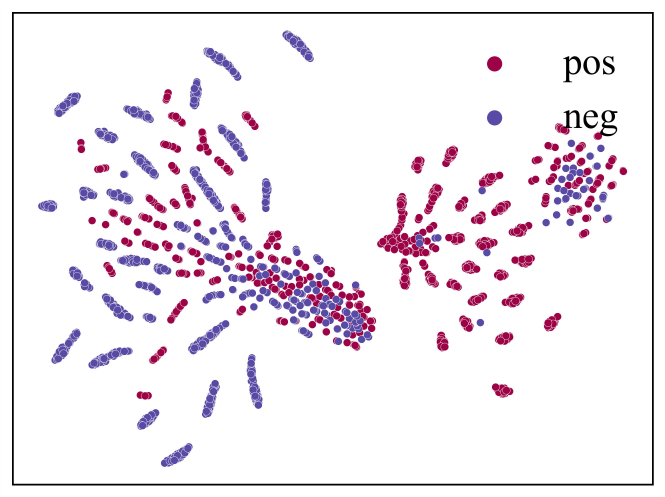}
			\end{minipage}
		}
		\caption{For every subfigure: the first row denotes the internal visualization of GMT in the S. cerevisiae datasets at $c = 0.3$; the second row denotes the internal visualization of Q-GAT in the \emph{S. cerevisiae} datasets at the perturbation level of $c = 0.3$. } 
  \vspace{-0.3cm}
  \label{fig4} 
	\end{figure} 

Table \ref{tbl6} shows that the prediction effect of Q-GAT (no GQT) and Q-GAT (no GQP) declines compared with that of Q-GAT on \emph{S. cerevisiae} datasets under different perturbation levels. Q-GAT (no GQT) has dropped most drastically. It means that both GQT and GQP have a positive effect on the robustness. 
Furthermore, one can see that when removing GQT, the performance of our model declines significantly on the \emph{S. cerevisiae} dataset. When removing GQP, the model's performance also decreases on \emph{S. cerevisiae} dataset, but not as severely as removing GQT. Therefore, GQT is more important to the model than GQP.

 \section{Robustness Analysis}

 \subsection{The SNR Analysis}

Now let us elucidate why the proposed Q-GAT can enjoy higher predictive performance and stability with respect to the noise perturbation from the perspective of the signal-to-noise ratio (SNR). The SNR is an important measure of the quality of signals, which is defined as the signal power divided by the noise power in the same area. Our overarching opinion is that the reason why a quadratic network can predict more accurately and achieve higher robustness to noise over other models, particularly models using conventional neurons, is because the nonlinearity embedded in quadratic neurons can effectively improve the SNR in the propagation of signals across layers. Specifically, in each layer, the quadratic operation can amplify the signal and inhibit the noise. What’s more, the cascade effect of the layer composition strengthens such an effect. When the SNR is enhanced, there allows more useful information and less unwanted noise, thereby facilitating the extraction of key features and the recognition of different classes of features. Thus, the decision-making process of a network is more accurate and robust.

Neurons are the basic components of neural networks, including graph neural networks. The representation ability of neurons to a large extent determines the overall performance of the network to a certain extent. Compared with linear neurons in traditional GNNs, the highlight of quadratic neurons is the nonlinearity in the simplified quadratic function. Chapeau-Blondeau \textsl{et al.} \cite{2006Nonlinear} proved that generically, a nonlinear device can amplify the SNR compared with linear filters. In the same vein, now let us show that the processing of quadratic neurons can result in a higher SNR than that of linear ones in the light of signal recovery.
    
Given the input signal $x(t) = s(t)+n(t)$, where $s(t)$ denotes the pure signal and $n(t)$ denotes the noise. The power of the pure signal $s(t)$ in $x(t)$ can be expressed as
	\begin{equation}
		P_{x_{s}}=\frac{1}{T} \int_{0}^{T}|s(t)|^{2} d t,
		\label{eq14}
	\end{equation}
	and the power of noise $n(t)$ in $x(t)$ is expressed as
	\begin{equation}
		P_{x_{n}}=\frac{1}{T} \int_{0}^{T}|n(t)|^{2} d t.
		\label{eq15}
	\end{equation}

Generally, the SNR is defined as the ratio between the power of signal and noise. To avoid negative values, the following Eq. \eqref{eq16} is adopted for the SNR of the input:
	\begin{equation}
		\mathcal{R}_{in}=10 \log \left(1+\frac{P_{x_{s}}}{P_{x_{n}}}\right).
		\label{eq16}
	\end{equation}
Now suppose that the signal $x(t)$ is fed into a quadratic network $Q$ to recover the pure signal:
	\begin{equation}
		y(t)=Q[s(t)+n(t)].
		\label{eq13}
	\end{equation}
For the signal recovery task, we can rewrite $y(t)$ as 
	\begin{equation}
	\begin{split}
		y(t)&=Q[s(t)+n(t)] \\
  &=s(t)+Q[s(t)+n(t)]-s(t)\\
		&=s(t)+n^{\prime}(t),
		\label{eq17}
	\end{split}
	\end{equation}
where $n^{\prime}(t)$ is regarded as the generalized noise. The power of pure signal $s(t)$ and noise $n^{\prime}(t)$ in $y(t)$ can be expressed, respectively, as Eqs. \eqref{eq18} and \eqref{eq19}.
	\begin{equation}
		P_{y_{s}}=\frac{1}{T} \int_{0}^{T}|s(t)|^{2} d t
		\label{eq18}
	\end{equation}
	\begin{equation}
		P_{y_{n}}=\frac{1}{T} \int_{0}^{T}|Q[s(t)+n(t)]-s(t)|^{2} d t
		\label{eq19}
	\end{equation}
	
Accordingly, the SNR of the signal being processed by a quadratic network is as follows:
	\begin{equation}
		\mathcal{R}_{Q}=10 \log \left(1+\frac{P_{y_{s}}}{P_{y_{n}}}\right)
		\label{eq20}
	\end{equation}
	
As shown in Eq. \eqref{eq21}, $V_{Q}$ represents the ratio of $\mathcal{R}_{Q}$ to $\mathcal{R}_{in}$, and $V_{Q}$ \textgreater 1 indicates an increase in SNR of the input signal.
\begin{equation}
		V_{Q}=\frac{\mathcal{R}_{Q }}{\mathcal{R}_{in }}=\frac{\log \left(1+\frac{P_{y_{s}}}{P_{y_{n}}}\right)}{\log \left(1+\frac{P_{x_{s}}}{P_{x_{n}}}\right)}
		\label{eq21}
	\end{equation}
	
Suppose that one uses a conventional network to perform the signal recovery, we have
	\begin{equation}
		z(t)=L[s(t)+n(t)].
		\label{eq24}
	\end{equation}
	Its ratio $V_{L}$ is expressed as
	\begin{equation}
		V_{L}=\frac{\mathcal{R}_{L }}{\mathcal{R}_{in} }=\frac{\log \left(1+\frac{P_{z_{s}}}{P_{z_{n}}}\right)}{\log \left(1+\frac{P_{x_{s}}}{P_{x_{n}}}\right)}.
		\label{eq25}
	\end{equation}
	
Now, we train a fully-connected one-hidden-layer conventional network (NN) and quadratic network (QNN), respectively, to compare their signal recovery effect from the perspective of the SNR. We set $s(t)=Acos(\omega t+\varphi)$, where $A=3$ to make sure $s(t)$ is not drowned in noise, $\omega=2\pi$, and $\varphi=0$. The noise $n(t)$ is Gaussian noise with different variance $\sigma^{2}$ ranging from 0.2 to 1. The algorithm is trained for 1000 epochs by Adam with an initial learning rate of 0.001. 


\begin{table}[htbp]
\setlength{\abovecaptionskip}{0.5em}
\setlength{\belowcaptionskip}{-0.5em}
\vspace{-0.2em}
\centering
\caption{The ratio of the output SNR to input SNR of NNs and a QNN with different layers in certain noise variance $\sigma^{2}$}
\scalebox{0.8}{
\begin{tabular}{|l|c|c|c|c|c|}
\hline
$\sigma^{2}$ & $0.2$  & $0.4$  & $0.6$ & $0.8$ & $1$        \\ \hline
$V_{L}$(NN-1) & 1.0210 & 0.6860 & 0.5523 & 0.7278 & 1.1925  \\ \hline
$V_{L}$(NN-5) & 1.0239 & 1.0427 & 1.1045 & 1.1342 & 1.1973  \\ \hline
$V_{L}$(NN-25) & 0.2107 & 0.2821 & 0.3538 & 0.3831 & 0.4135  \\ \hline
$V_{Q}$(QNN-1) & \textbf{1.0240} & \textbf{1.0449} & \textbf{1.1055} & \textbf{1.1359} & \textbf{1.1984}  \\ \hline
\end{tabular}}
\label{tab:CNN_QNN_comparison}
\vspace{-1em}
\end{table}

\begin{figure}[htbp]  
  \setlength{\abovecaptionskip}{0.5em}
\setlength{\belowcaptionskip}{0em}
  \vspace{-0.2cm}
    \centering
    \includegraphics[width=\linewidth]{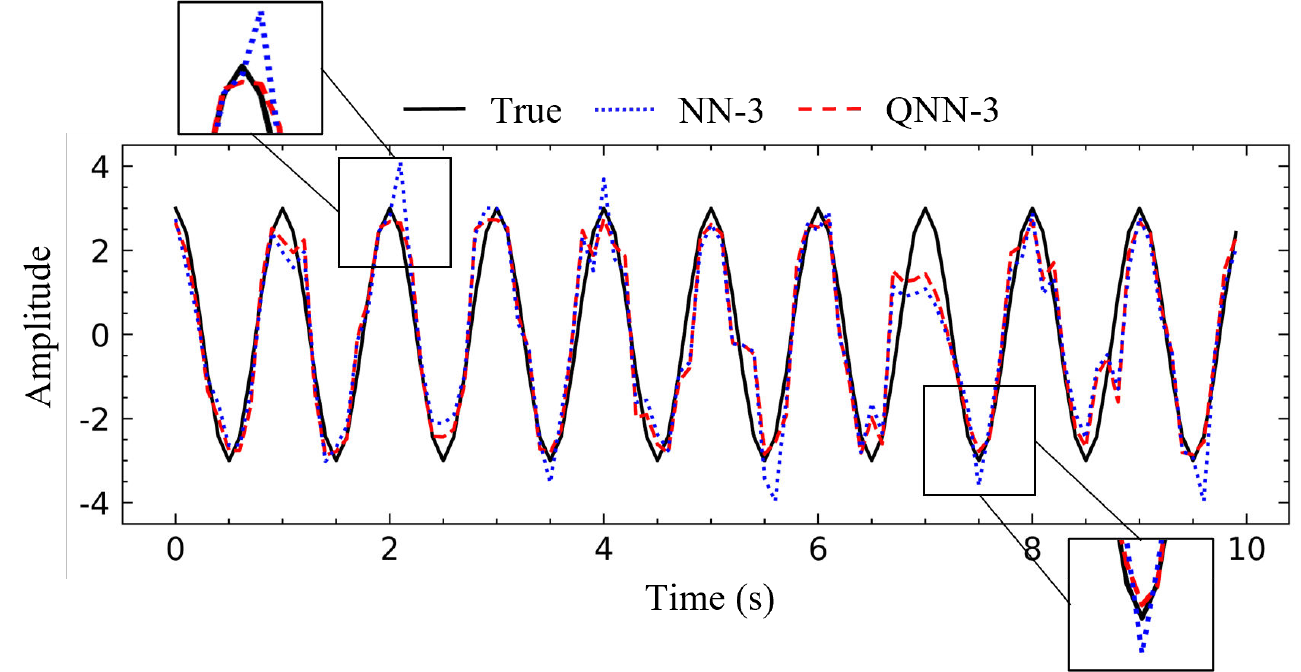} 
    \caption{A comparison between reconstructed signals of a NN and a QNN of 3 layers.} \label{figm}
    \vspace{-1em}
    \end{figure}

\Cref{tab:CNN_QNN_comparison} contrasts $V_{L}$ and $V_{Q}$ of NNs and a QNN with different layers under different noise variances, where the number of neurons is 5 in every intermediate layer of NN-5 and NN-25. $V_{L}$ of NN-1 is far lower than 1 at certain noise levels \textit{e.g.}, $\sigma^2=0.4,0.6$, which means that the conventional NN can hardly work. However, $V_{Q}$ are basically always greater than 1 at every noise level. As the noise level goes up, $V_{Q}$ shows an overall upward trend. It indicates that quadratic neurons can inhibit the noise and boost the SNR ratio markedly. However, no matter how many layers, the NN can not improve SNR better than QNN-1, even when the number of layers reaches 25. Figure \ref{figm} shows the reconstructed signals of NN-3 and QNN-3. QNN-3 reconstructs the periodicity and amplitude of the signal well, but NN-3 almost fails. Based on this synthetic experiment, we think that quadratic neurons are more helpful in suppressing the impact of noise.

 \subsection{Interpretability Analysis}

Here, we dissect the robustness of the proposed Q-GAT from the perspective of interpretability. Interpretability and robustness are two major weaknesses of a deep learning model. The former is that predictions of deep learning are hard to explain, while the latter is that the predictions of deep learning are easily changed by small adversarial samples. Interpretability and robustness are highly related. For example, saliency methods \cite{fan2021interpretability} that use gradients of a deep model with respect to the input to explain where a model deems as important are the mainstream methods in interpretability studies. Interestingly, such gradients are also used to generate adversarial examples. 

We argue that interpretability studies can shed light on robustness by checking if the model leverages the correct features to predict. For example, a picture with a polar bear may also contain snow or ice, the model may misuse the information about snow or ice to classify the polar bear rather than the features of polar bears. We think \textit{if a model can use as many effective features as possible, it is more robust to noise perturbation.} Along this direction, as we know, Q-GAT is obtained by substituting conventional neurons with quadratic neurons in GAT. While GAT can assign different weights to neighbors of each node as their importance scores, it ignores the importance of features of each node. However, gene features in a GRN also contain abundant important information, which should be mined and interpreted appropriately. Compared with GAT, Q-GAT can additionally assign weights for features of nodes because of the \textit{qttention} mechanism in quadratic neurons, therefore facilitating the use of more feature information.

\begin{figure}[htbp]  
  \setlength{\abovecaptionskip}{0.0em}
\setlength{\belowcaptionskip}{0em}
  \vspace{-0.3cm}
    \centering
    \includegraphics[width=0.46\textwidth]{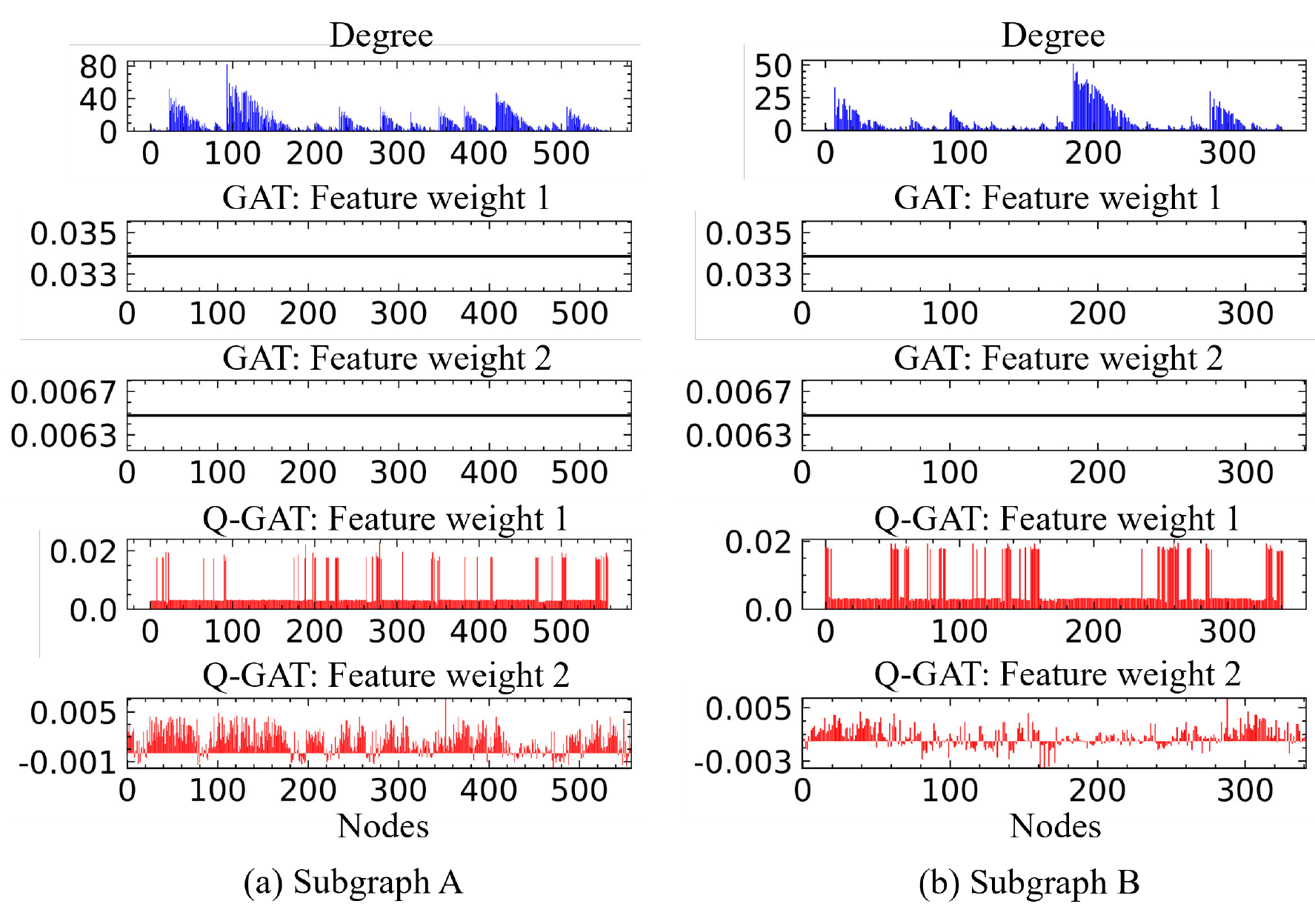} 
    \caption{The node degree and feature weights of GAT and Q-GAT in subgraphs A and B} \label{figl}
    \vspace{-0.4cm}
    \end{figure}

Figure \ref{figl} shows the gene nodes on the horizontal axis and feature weights for each node on the vertical axis in a randomly selected subgraph of the \emph{E. coli} dataset. GAT always assigns the same weight to each node feature even in different subgraphs. However, Q-GAT extracts some key nodes by assigning different feature weights, \textit{i.e.}, some features are favored for link prediction, while some are not. This means that the gene features are also leveraged in link prediction. Therefore, Q-GAT is a multi-dimensional graph attention model from feature to node and node to node. In contrast, GAT is just a single-level attention model from node to node.

Next, we investigate whether the key nodes mined by Q-GAT are really critical or not. The node degree often serves an indicator to measure the importance of nodes which reflects the connectivity of edges. The nodes with a high degree are generally considered important. As shown in the bottom row of \Cref{figl}, we plot the node degree of the subgraph. But the key nodes extracted by Q-GAT and the node degree are not very close. 

Therefore, we computationally verify the importance of key nodes extracted by Q-GAT. We set the nodes with feature weights greater than 0.0175 to 0 in subgraph A of entire graph, and we set the same number of nodes with the highest node degree to 0 in the graph. Then, we compare the results with overall predictions for the entire graph. \Cref{tab:comparison} shows that when we suppress the effect of key nodes extracted by Q-GAT, the accuracy of prediction in Q-GAT is compromised. However, the overall prediction accuracy does not change a lot in Q-GAT when the role of high-degree nodes is suppressed. We also test the importance of key nodes extracted by Q-GAT in GAT model. The same phenomenon can be found. These results suggest that Q-GAT can extract important nodes more accurately than node degree.

To summarize, our interpretability analysis reveals that Q-GAT can leverage both feature importance and node importance in predictions. The employment of more useful information endows Q-GAT with a more powerful ability to confront adversarial perturbation.

\begin{table}[htbp]
\setlength{\abovecaptionskip}{0.5em}
\setlength{\belowcaptionskip}{-0.5em}
\vspace{-0.2em}
\centering
\caption{Prediction results obtained by inhibiting key genes determined by qttention weights or node degree in subgraph A at the perturbation level of c = 0.1.}
\scalebox{0.8}{
\begin{tabular}{|l|c|c|}
\hline
& Q-GAT  & GAT               \\ \hline
Original & 0.7265 & 0.6997              \\ \hline
qttention weights & 0.7037 & 0.6810  \\ \hline
node degree & 0.7224 & 0.7037      \\ \hline
\end{tabular}}
\label{tab:comparison}
\vspace{-1em}
\end{table}

\section{Conclusion}    

In this paper, we have proposed Q-GAT for the construction of GRNs. To assess the stability of Q-GAT, we have added various levels of perturbation to gene expression data and then tested different models' inference performance in the \emph{E. coli} and \emph{S. cerevisiae} datasets. The comparative experiments suggest that Q-GAT is more stable than the start-of-the-art methods under different perturbation levels. Furthermore, we have analyzed why Q-GAT is more robust from the perspective of signal processing and interpretability. In the future, Q-GAT can be upgraded by combining other advanced techniques to further improve its robustness.

 \section*{Acknowledgement}
 This work was supported by  the Natural Science Foundation of Liaoning Province (2022-MS-114).


\bibliographystyle{unsrt} 
\bibliography{references.bib}


\end{document}